\documentclass{ws-procs975x65}

\def\be{\begin{equation}}
\def\ee{\end{equation}}
\def\ben{\begin{eqnarray}}
\def\een{\end{eqnarray}}
\newcommand{\bec}{\begin{center}}
\newcommand{\eec}{\end{center}}

\def\ba{\begin{eqnarray}}
\def\ea{\end{eqnarray}}

\def\negenspace{\kern-1.1em}

\def \quabla{\sqcup \kern-9pt \sqcap}
\begin{document}

\title{DARK MATTER HALOS AS BOSE-EINSTEIN CONDENSATES}

\author{ECKEHARD W. MIELKE}

\address{
Departamento de F\'{\i}sica,\\
Universidad Aut\'onoma Metropolitana-Iztapalapa,\\
Apartado Postal 55--534, C.P. 09340, M\'exico, D.F., MEXICO\\
E-mail address: ekke@xanum.uam.mx}

\author{BURKHARD FUCHS}

\address{
Astronomisches Rechen-Institut, M\"onchhofstr.~12--14, 69120
Heidelberg, Germany\\
E-mail address: fuchs@ari.uni-heidelberg.de}

\author{FRANZ E. SCHUNCK}

\address{
Institut f\"ur Theoretische Physik, Universit\"at zu K\"oln,
50923 K\"oln, Germany\\
E-mail address: fs@thp.uni-koeln.de}

\maketitle

\abstracts{
Galactic {\em dark matter} is modelled by a scalar field
in order to {\em effectively} modify Kepler's law {\em without} 
changing standard Newtonian gravity.
In particular,  a {\em solvable} toy model with a
{\em self-interaction} $U(\Phi)$ borrowed
from non-topological solitons produces already qualitatively correct
{\em rotation curves} and {\em scaling relations}. Although relativistic 
effects in the halo are very
small, we indicate corrections arising from the general
relativistic formulation. Thereby, we can also probe  
the weak gravitational lensing of our soliton type
halo. For cold scalar fields, it corresponds to a gravitationally confined 
Boson-Einstein condensate, but of galactic dimensions.\\
\\
%KEYWORDS: Dark matter,  rotation curves,  scaling relations, scalar field model, Bose-Einstein condensate.
}

%*********************************
\section{Introduction}

The existence of {\em dark matter} was postulated for the first time by
Zwicky\cite{zwi} in 1933. He showed that the virial masses of clusters of
galaxies exceed the mass of luminous matter by roughly one order of magnitude.
In the 1970's spectrographs became sensitive enough that rotation curves of
spiral galaxies could be observed in large numbers by Rubin and
collaborators\cite{RFT80,RBFT85}. They all showed a systematic tendency to
stay flat as function of galactocentric radius,
$v_\varphi (r) \simeq {\rm const.}$,
whereas the mass distributions of the bulges and the radially exponentially
declining surface densities of the galactic disks  lead to the prediction of
falling rotation curves. Evidence for an extra mass component became even more
decisive when HI--rotation curves\cite{ABBS} were observed, because the
neutral hydrogen gas extends radially outwards many disk scale lengths. Even
though there is a long debate\cite{Bo03} to what degree the inner parts of
galaxies are dominated by dark matter, the flat outer HI-rotation curves prove
beyond doubt the existence of dark matter halos. The analysis of the kinematics
of satellite galaxies\cite{zar2} showed that the dark halos extend out to
several hundred kpc with the cumulative mass still rising linearly with
galactocentric radius, $M(r) \propto r$, which corresponds to flat rotation
curves.

Persic {\em et al.}\cite{PS96} investigated statistical correlations between
structural parameters of visible and dark matter using a data set of
almost 1000
galaxies and developed the model of the so called `universal rotation curve'.
Accordingly, galaxies of low luminosity are more dominated by dark matter than
bright galaxies. Dwarf irregular galaxies and low surface brightness galaxies
seem to be dominated completely by dark matter. Burkert\cite{Bu95,BS99} finds
for such galaxies an empirical fit to the rotation curves which implies
asymptotically $v_\varphi^2 \rightarrow \ln r/r$, i.e.
a logarithmic modification of the Kepler law.

For several classes of gravitational theories, it has been argued
that the introduction of dark matter is necessary\cite{nester}, although
in the modified Newtonian dynamics (MOND) scheme,
exact expressions were derived
for the Newtonian potential and for the rotation curve\cite{BM95}. However, recent 
observational constraints\cite{AB01} make it more and more challenging to interpret 
galactic rotation in terms of new gravitational physics\cite{Be88}.

An intriguing possibility is a {\em scalar field} model of the
halo first proposed by
one of us\cite{Sch94,schunck,sch97,Sc98,Sc99} using a
complex massless field coupled to the Einstein
equation, an idea which was later taken up by several other authors.
Hence, the complex massless scalar field forming star-like
galactic halos embodies another form of {\em dark energy}.

Massive compact halo objects, so-called MACHOs\cite{alc},
consisting of baryonic matter or, alternatively, of {\em boson star}s
(BSs)\cite{MS00,MS02a,SM03} do not seem to be
sufficient to resolve this problem completely.
More recent approaches\cite{SKM03,FM04} propose
connections between scalar field matter and the cosmological
constant.

On the other hand, the standard {\em cold dark matter} (CDM) model, where dark
matter particles interact only
through gravity, may be in conflict with observational features
on Mpc scale. In particular,  the
CDM model of Navarro, Frenk and  White (NFW)\cite{b1} 
tends to predict density profiles of dark halos
which are cuspy at the center\cite{a17}.
Therefore, Spergel and Steinhardt\cite{{SS00}}
proposed that dark matter is self-interacting, an idea which found ramification in
 Refs.~\refcite{RT00,a13}. The evolution of the {\em self-interacting} 
 dark matter  has been recently analyzed in
more detail\cite{BS02,BSI02}.

In the same vein,  we consider here a model\cite{MS02,MSP02} where dark matter is modelled by
a primordial scalar field with a {\em self-interaction} $U(\Phi)$.
Since the observed rotation
velocities are roughly bounded\footnote{The maximum velocity within
a rotation curve of almost
400 km/s is found in a Sa galaxy\cite{RFT80}; the velocity of 367 km/s
at the farest  measured
point  belongs also to a Sa galaxy\cite{RBFT85}.}  by
$v_\varphi \leq 10^{-3}c$, {\em i.e.} are non-relativistic,
a Newtonian type approximation sufficies.

The formation of dark halos as a gigantic Bose-Einstein condensation
in early Universe is discussed in Sec.~\ref{BEC}.
In particular, a boson star like toy model
with a $\Phi^6$ repulsive self-interaction is known\cite{Mi78} to allow
in the limiting case of {\em flat} spacetime  an exact spherically
symmetric {\em soliton} solution of the corresponding
nonlinear Klein-Gordon equation, as is shown in Sec.~\ref{ntssol}.
Simulating the halo by such a {\em non-topological soliton} (NTS) for
the {\em positive} range of the potential,
yields a Newtonian mass distribution which provides a qualitatively
rather good fit to the
rotation curve data of dwarf irregular and low surface brightness
galaxies,
see Sec.~\ref{rotcur} and the comparison with observations
in Sec.~\ref{observ}.
In Sec.~\ref{outlook}, we indicate the extension to the general
relativistic
formulation and speculated on the superstring nature of the dark matter
scalar.
In Sec.~\ref{lensing}, we display the gravitational lensing properties
of the NTS halo.

%*********************************
\section{Dark halo as a self-gravitating Bose-Einstein condensate? \label{BEC}}
Since Einstein and Bose it is well-known that
identical integer spin particles
can occupy the same ground state. Such a {\em Bose-Einstein condensate}
(BEC) has been experimentally realized in 1995 for cold atoms of even
number of electrons, protons, and neutrons, see Anglin and
Ketterle\cite{AK02} for a  review. In the mean-field ansatz, the
interaction of the atoms in a dilute gas is approximated by
the effective potential
\be
U(|\Psi|)_{\rm eff}= \frac{\lambda}{4} |\Psi|^4 \, .  \label{Ueff}
\ee
This leads to a {\em nonlinear} Schr\"odinger equation
for the macroscopic matter wave field $\Psi$,
in this context known as Gross-Pitaevskii equation.
In a microscopic approach, one introduces  bosonic creation and
annihilation operators $b^\dagger$ and $b$, respectively, satisfying
$[b,b^\dagger]  =1 $,
and finds that every number conserving normal ordered correlation
function
$\langle b_1 \cdots b_n \rangle$ splits into the sum of all possible
products of contractions $\langle b_i  b_j \rangle$ as in
Wick's theorem of quantum field theory (QFT). For $n=1$
one recovers the Gross-Pitaevskii equation, whereas the next order leads
to  Hartree-Fock-Bogoliubov equations\cite{KB01}.

In the context of astrophysics, an assembly of massive  real scalar particles
has  already been considered  by Bonazzola and Pacini\cite{BP66} in 1966.
Later on, solutions for ``systems of
self-gravitating'' bosons were considered by Ruffini
and Bonazzola\cite{RB69} similarly to the Hartree-Fock atom.
Because a real scalar has no {\em anti-particle} states,
the corresponding Klein-Gordon field $\Phi$ 
can only be   decomposed into positive and negative
frequency field operators:
\be
\Phi=\Phi^+ + \Phi^- \, .
\ee

For  a spherically
symmetric configuration
\ba
\Phi^+ & = & \sum_{nla} b_{nla}^\dagger R_l^n(r)
 Y^{|a|}_l(\theta, \, \varphi)
  e^{-i \omega_{nl} t}\,,\nonumber \\
  \Phi^- & = & \sum_{nla} b_{nla} R_l^n(r)
 [Y^{|a|}_l(\theta, \, \varphi)]^*
  e^{+i \omega_{nl} t}
\ea
can be   expanded in terms  of the wave function
of a {\em bound  state} similarly as for the hydrogen atom.
Here $b_{nla}$ and $b_{n'l'a'}^\dagger$ are again creation
and annihilation operators,
$R_l^n(r)$ are radial distributions,
$Y^{|a|}_l(\theta, \, \varphi)=(1/\sqrt{4\pi})
P^{|a|}_l(\cos\theta)\, e^{-i|a|\varphi}$ the spherical harmonics
given in terms of the
normalized Legendre polynomials, and $|a|\leq l$ are  the quantum numbers
of {\em azimuthal}  and {\em angular} momentum. (The nodeless radial field
solution $R_{0}^0(r)$ will later be denoted by $P(r)$.)

For the ground state of a cold configuration,
\be
|N\rangle=|N,0,0,\ldots\rangle:=\prod_1^N b_{100}^\dagger \vert0\rangle
\ee
is chosen\cite{RB69},
where $\vert0\rangle$ is a vacuum state in the curved spacetime
`background'.
The energy-momentum tensor $T_{\mu \nu}(\Phi)$ of the scalar
field becomes now an operator. When coupled to gravity,
 the vacuum expectation value
$\langle T_{\mu \nu}\rangle:= \langle N|:T_{\mu \nu}:|N\rangle$
for the ground state is inserted  into  the right-hand side of the Einstein equation
(\ref{phi152}), where
$:T_{\mu \nu}:$ denotes normal ordering of the operator products.

An {\em excited state}  $\vert N,n,l,a\rangle:= \Phi^+ \vert0\rangle$
of positive energy  can be obtained by applying creation operators.
Such a `gravitational atom'\cite{FG89} represents a {\em coherent}
quantum state, which nevertheless can have macroscopic size and large
mass.
The gravitational field is self-generated
via the mean value $\langle T_{\mu \nu}\rangle$ of the energy-momentum tensor, but remains
completely classical, whereas the real scalar is treated to some
extent as operator.

Below some  critical temperature (\ref{critT}), such
configurations of coherent
bosons would form  Bose-Einstein condensates on an astrophysical scale\cite{BB02}, 
and following T.D.~Lee\cite{lee} are
called boson stars (BSs).

Real scalars have the disadvantage that no local symmetry provides a
conserved particle number,
and so one needs to
introduce by hand a normalization condition in order to stabilize the system.

It is gratifying to note that {\em BSs} with
{\em repulsive} self-interaction $U(|\Phi|)$ considered
already by Mielke and Scherzer\cite{MS81} and Colpi {\em et al.}\cite{CSW86}
have their counterparts in the
effective potential (\ref{Ueff}) of BEC.
Thus  a cold BS can be considered as a {\em self-gravitating} BEC on an
astrophysical scale\cite{HBG00,JB01,BLV01}.
Moreover, BSs of rather different sizes can occur:
it could be just a `gravitational atom'; it could be as massive
as the presumed black hole in the central part
of a galaxy; or it could be an alternative explanation
for parts of the dark matter in the halo of galaxies, as we are
going to explain
below. BSs, if they exist, would be an astrophysical realization with a
self-generated gravitational confinement.

The idea of a cold mini-boson star as a BEC condensate on a galactic scale
was reiterated, without referring to the earlier papers,
in Refs.~\refcite{HBG00,HBG03} in a Newtonian approximation.
A comparison of the Jeans scale of
a dark matter halo  and the de Broglie wave length of the gravitationally
confined bosons provides the  estimate $m_\Phi \simeq 10^{-22}$ eV/c$^2$
for a tiny `bare' mass of the bosons.

Bosonic particles such as fundamental scalars allow the possibility of
a self-gravitating BEC on
an astrophysical scale, commonly referred to as BS.
However, a repulsive self-interaction
has the physical effect of  delaying the progressive
collapse of a cosmic assembly formed by accretion.
This, in particular, applies to a {\em cold} BS below the
critical temperature
\be
T_{\rm c}= \frac{2\pi \hbar^2}{mk}\left[\frac{N}{\xi(3/2)V}\right] \, ,  \label{critT}
\ee
where $\xi(z)$ is the Riemann zeta function, provided
an ideal gas approximation is employed. However,
for the possible formation of a BEC as a dark halo,
one has to take into account its immersion into the heat bath of
the cosmic background radiation  of nowadays
$2.726$ K.

The physical nature of gravitationally coupled scalar fields\cite{Br96} 
is at best speculative:
{}From the field-theoretical
point of view, a {\em dilaton field} $\varphi$ coupled universally to the
trace of the energy-momentum tensor via
\be
L_{\rm int}:= \frac{1}{M_{\rm Pl}}\varphi  T^\mu_\mu  \label{trace}
\ee
is a rather natural candidate. In order to apply this to galactic halos,
one needs an ultralight
scalar field with a Compton wave length $\lambda= \hbar/2m$ of the
order of the galactic core
$r_{\rm c}\simeq 10^{20}$ m, corresponding to  $m_\varphi\simeq 10^{-22}$ eV/c$^2$.
On the other hand, ultra-light `gravi-scalars' arising 
in the framework of five-dimensional 
braneworlds are constrained\cite{DK03} by the precisely observed slow-down 
of the Hulse-Taylor binary pulsar due the gravitational wave emission.
The universal coupling (\ref{trace}) provides a natural mechanism\cite{DGS01} to protect
such a tiny mass from renormalization due to  quantum  loops arising in the Standard Model.
In order to suppress large long-range forces, an approximate global symmetry needs to be 
postulated which, nevertheless,  allows a coupling of $\varphi$ to the 
Pontryagin term of electromagnetism via  $(\varphi/M_*) F\wedge F$, 
{\em cf.}  Carroll\cite{Ca98}.

A related proposal is that of Wetterich\cite{We01} which endows a
massless  dilaton (as a pseudo-Goldstone boson)
with a tiny mass via the conformal anomaly.  Such a dynamical scalar
``quintessence" may also account
for {\em dark energy}\cite{BB03}.
Another approach\cite{PC02} to unify clustered dark matter
and dark energy is based on a Born-Infeld type higher derivative
Lagrangian for the same hypothetical scalar field.

Recently\cite{SKM03}, we analyzed the {\em bifurcations} in the mapping
of a self-interacting scalar field $\Phi$ minimally coupled to
Einstein's general relativity (GR) to
a non-linear curvature scalar Lagrangian.
Intriguingly, the higher-order Lagrangian bifurcates in almost
linear Einsteinian patches, but with different effective gravitational
strength and cosmological constant (dark energy)
depending on the cosmological scale.
More precise constraints on dark energy are expected to come 
from future data for the cosmic microwave background
(CMB) radiation\cite{BB03}, 
which may also get modified\cite{MS04} via  the 
Sunyaev-Zeldovich effect  during its long journey. 

%*********************************
\section{Newtonian non-topological solitons \label{ntssol}}
As a solvable toy model with {\em self-interaction}, let us consider the 
Klein-Gordon equation (\ref{phi153}) with
a $\Phi^6$ type potential
\be
U(\vert  \Phi \vert ) = m^2 \vert  \Phi \vert ^2 \left(1 -
\chi \vert  \Phi \vert^4 \right), \qquad \chi \vert  \Phi \vert^4\leq 1\,,
\label{pot6}
\ee
where $m$ is the `bare' mass of the boson and  $\chi$
a coupling constant, which are thought of as constants of nature.
The self-interaction in the
radial Klein-Gordon equation  takes the form
$ dU(P)/dP^2=  m^2  - 3 m^2 \chi P^4$. In flat spacetime, such a
model  was first considered by one of us\cite{Mi78} for constructing
{\em non-topological  soliton} (NTS) solutions.

For a spherically symmetric configuration and
 the choice $\omega =m$, the corresponding
nonlinear Klein-Gordon equation simplifies  to an {\em Emden type
equation}
\be
P^{\prime\prime} + \frac{2}{x} P^{\prime}
  + 3\chi P^5 = 0\, ,
\label{Emden}
\ee
familiar from the
astrophysics of gaseous spheres. It has
the completely {\em regular} exact solution
\be
P(r) = \pm \chi^{-1/4} \sqrt{\frac{A }{1+A^2 x^2}} \, , \label{NTS}
\ee
where  we introduced
the {\em dimensionless} radial coordinate $x:=m r$,
and $A = \sqrt{\chi} P^2(0)$ in terms of the central value.
The solution depends
essentially on the nonlinear coupling parameter $\chi$, since the limit
$\chi\rightarrow 0$ would be singular. This feature is rather
characteristic  for {\em soliton
solutions}.  Already 1978,  it has been generalized\cite{Mi78}
to a NTS with (quantized) angular momentum $l$.

In the following, it suffices to restrict
ourselves to the above given range  for which
the potential $U(\vert  \Phi \vert)$  remains {\em positive}.
If effects of
self-gravitation are taken into account, this scalar potential needs
not to be bounded from below: In the case of
general relativistic BSs\cite{MS81,kusmartsev,KMS91,SKM92,MS96,SM98,MS01}
with a self-interaction $\lambda |\Phi|^4$,
it has been proven numerically\cite{SL97} that NTS exist even for
negative values of $\lambda$.
The extension to {\em soliton stars}\footnote{Our potential (\ref{pot6})
is not bounded from below,
however, its positive range is sufficient for applying our toy model.
An example of a bounded potential
is $U(\vert  \Phi \vert ) = m^2 \vert  \Phi \vert ^2 \left(1 -
\ \vert  \Phi \vert^2/\vert  \Phi_0 \vert^2\right)^2$
typical for soliton stars,
but then simple analytical expressions are not available and one
needs, for large radius, to deal
with asymptotic solutions like
$P(r) \rightarrow \pm \sqrt{A/(1+B^2 e^{2x})} $.}
have been considered by Lee and Pang\cite{lee}  as well as Gleiser\cite{GL89}.

The canonical energy-momentum tensor of a relativistic spherically
symmetric
scalar field is diagonal,
{\em i.e.}~$T_\mu{} ^\nu(\Phi) = {\rm diag} \; (\rho , -p_r,$
$-p_\bot, -p_\bot )$ with
\ba
\rho &=& \frac{1}{2} \left ( \omega^2 P^2  +P'^2  + U
 \right ) \; , \nonumber  \\
p_{\rm r} &=&  \rho -  U \; , \;  \nonumber  \\
p_\bot &=&  p_r -  P'^2  \; ,    \label{emt}
\ea
where $'=d/dr$. The form (\ref{emt}) is familiar from perfect fluids,
{\em except} that the radial and tangential pressure generated by the
scalar field are in general different, {\em i.e.}~$p_r \neq p_\bot $.
This {\em anisotropy} of scalar matter should not
 been ignored, since it holds even in
flat spacetime, or in the Newtonian approximation.

{}From (\ref{emt}) we find in flat spacetime the energy-density
\ba
\rho &=& \frac{m^2}{2} \left [ 2 P^2
 + P'^2 - \chi P^6 \right ]  \nonumber \\
 & = & \frac{A m^2}{2\sqrt{\chi}(1+A ^2 x^2)}\left[2
 +\frac{ A ^4 x^2 -A ^2}{(1+A ^2 x^2)^2}  \right ]  \; .
\label{rhonts}
\ea
(If we would consider a real scalar field instead, there is merely
the change of the first proportionality constant  2 in the bracket to 1.)
The requirement of the positivity  of $U(\Phi)$ at the
origin yields the  constraint  $A < 1$, which we will adopt in the
following.

The leading term of the Newtonian type mass concentration (\ref{rhonts})
is exactly the density law of the {\em quasi--isothermal sphere}\cite{IS00}
\begin{equation}
\rho (r) \simeq
\frac{\rho_0 r_{\rm c}^2}{r_{\rm c}^2+r^2}  \; .
\label{quasirho}
\end{equation}
At large radii the density falls of like $\rho \propto r^{-2}$ which
corresponds to an asymptotically flat rotation curve. Comparing  with the
quasi--isothermal sphere (\ref{quasirho}), the central density of 
 the NTS model is given by
$\rho_0 \simeq A  m^2/\sqrt{\chi}$ and the core radius is $r_{\rm c}\simeq
1/m A$. This implies a {\em scaling law} for the dark halos of the form
\begin{equation}
\rho_0 \simeq \frac{m}{\sqrt{\chi}}\frac{1}{r_{\rm c}}\propto \frac{1}
{r_{\rm c}} \; ,  \label{scallaw}
\end{equation}
where $A$, which may vary from halo to halo, {\em cancels out}.

This nonlinearly coupled scalar field excerts the
following radial and tangential pressures:
\ba
p_{\rm r} &=& \frac{m^2}{2} \left [
 \chi P^6 + P'^2  \right ] =
\frac{A ^3 m^2}{2\sqrt{\chi}(1+A ^2 x^2)^2} \simeq
  \frac{A ^3 m^2}{2\sqrt{\chi}} \; , \label{radp} \nonumber \\
  p_\bot  &=& \frac{m^2}{2} \left [
 \chi P^6 - P'^2  \right ] =
\frac{A ^3 m^2(1-A ^2 x^2)}{2\sqrt{\chi}(1+A ^2 x^2)^3} \simeq
  \frac{A ^3 m^2}{2\sqrt{\chi}} \; , \nonumber\\
  \rho + p_{\rm r}&=&  m^2 [ P^2 + P'^2 ] \nonumber\\
  &=&
  \frac{A m^2}{\sqrt{\chi}(1+A ^2 x^2)}\left [
   1 + \frac{A^4  x^2}{(1+A ^2 x^2)^2}  \right ] \; .
\ea
Thus, at the center of the NTS, we have $p_{\rm r}(0)= p_\bot(0)$.
Asymptotically, we find at radial infinity
\be
p_{\rm r}\, , - p_\bot \rightarrow \frac{m^2}{2\sqrt{\chi} A  x^4} \; .
 \label{radpa}
\ee

{}From the energy density (\ref{rhonts}) 
the  {\em mass function}
\be
M(r):= \int_0^r \rho y^2 dy  \label{massf}
\ee
can be obtained by straightforward integration.
With the aid of {\sc Reduce}\cite{Sch99} we find for our model:
\ba
M(r) &=& \frac{1}{m A \sqrt{\chi}}\left[x
 +\frac{A ^2 -8}{8A }\arctan(A  x)
 -\frac{A ^2 x}{8}\frac{1+3A ^2 x^2}{(1+A ^2 x^2)^2}
\right] \; , \nonumber \\
 &\simeq & \frac{A}{6 \sqrt{\chi }} m^2 \left ( 2 - A^2 \right )
    \cdot r^3 \; , \nonumber \\
 &\rightarrow & \frac{r}{A \sqrt{\chi}} \; .
\label{massfunc}
\ea

 %**********************************
\section{Rotation curves \label{rotcur}}

The tangential velocity $v_\varphi$ of stars moving like
``test particles" around the center of
a galaxy is not directly measurable, but can be inferred from the
redshift $z_\infty$ observed at spatial infinity, for which
\be
(1+z_\infty)^2 =\frac{e^{\nu_\infty}}{e^{\nu}}
 \frac{(1 \pm v_\varphi)^2}{1 - v^2_\varphi}
\ee
holds. Due to their non-relativistic velocities in galaxies bounded
approximately by
$v_\varphi/c \leq 10^{-3}$, we observe $z_\infty  \simeq v_\varphi$
(as first part of a geometric series)
with the consequence that the lapse function necessarily
tends to unity, i.e.~$e^\nu \simeq
e^{\nu_\infty}/(1 -v^2_\varphi)\rightarrow 1$.
In general, there are two dominating influences
on the redshift of  stars: The gravitational potential of all
matter components within the galaxy (stars, gas, dark matter) and the
Doppler effect. (Since for
dwarf galaxies dark matter dominates up to the center, we can
neglect the  contribution of stars and gas.)
If $e^{\nu_\infty}=1$, the redshift for receding stars is
$z=e^{-\nu /2} \sqrt{(1-v_\varphi )/(1+v_\varphi )}-1$.
For example\cite{Sc98}, the influence of the gravitational field
of the dark matter on the redshift is just about 1.5 km/s, hence,
negligible. Of course, if the gravitational potential has the same
order of magnitude as the Doppler effect, then a velocity of 300 km/s
which observations reveal so far, would be actually just 150 km/s,
but this would certainly demand too strong gravitational potentials
for galaxies.

In general, for the static spherically symmetric metric (\ref{met})
considered lateron, an observer at rest at the equator of the  Schwarzschild
type
coordinate system measures the following
tangential velocity squared as a point particle
(or a star, regarded as a sufficient small body\cite{EG04})
flies past him in its circular orbit,
{\em cf.} Misner {\em et al.}\cite{mtw}, p.~657, Eq.~(25.20):
\ba
v^2_\varphi &:=& e^{-\nu} r^2 \left ( \frac{d\varphi}{dt} \right )^2
= \frac{1}{2} r\nu^\prime
= \frac{1}{2}\left[1 -
 e^{-\lambda}  +\kappa p_{\rm r}r^2 \right]e^{\lambda }\nonumber\\
 &=& \frac{\kappa}{2}\left[\frac{M(r)}{r} + p_{\rm r}r^2\right]\exp\left\{
 \frac{r}{r-\kappa M(r)}\right\}
 \nonumber \\
 &\simeq& \frac{\kappa}{2}\left[\frac{M(r)}{r} + p_{\rm r}r^2\right]
 \, . \label{tanvel}
\ea

Outside of matter ($\nu = -\lambda$) for a weak gravitational field,
Eq. (\ref{tanvel}) reduces to the
Newtonian form $v_{\varphi,{\rm Newt}}^2 \simeq M(r)/r$.
As is well-known\cite{wald,SH03}, a naive application of the Newtonian limit
would have led us to geodesics in flat spacetime, i.e.~as if
gravity would not affect the motion of test bodies like our stars moving
in the dark matter halo. Thus it is mandatory to go beyond, as is
indicated by our approximation of the generally
relativistic formula (\ref{tanvel}).
Then,  also the  pressure component $p_r \neq 0$ of an anisotropic `fluid'
contributes, our prime example being the case of scalar
fields. In our NTS model, however, due to the fast decrease of $p_r$,
{\em cf.}~Eq.~(\ref{radpa}), its contribution to the asymptotic value of
the rotation velocity is almost negligible.

{}From the mass function
(\ref{massfunc}) of the Newtonian NTS solution (\ref{NTS}) and its radial
pressure (\ref{radp}), we find for the rotation velocity
\be
v^2_\varphi /v^2_\infty =
 1 +\left(\frac{A ^2}{8}-1\right)\frac{\arctan(A  x)}{A  x}
 + \frac{A ^2}{8}\frac{A ^2 x^2-1}{(1+A ^2 x^2)^2}
  \, , \label{NTSrot}
\ee
for which the following approximations
\ba
 v^2_\varphi &\simeq& \frac{\kappa A  (1+A ^2)m^2}{6\sqrt{\chi}} r^2
\nonumber\\
 v^2_\varphi &\rightarrow&  v^2_\infty = \frac{\kappa}{2A \sqrt{\chi}}=
 \frac{\kappa}{2\chi P^2(0)} \leq 10^{-6},
\ea
hold near the center and at the far field, respectively. Together with
(\ref{massfunc}) we conclude that asymptotically $M\rightarrow 2 v^2_\infty r$.

Observationally, there is the rough restriction $v_\varphi/c \leq
10^{-3}$ of the rotation velocities of galaxies, which
can be used to constrain the mass $m\leq 10^{-22}$ eV/c$^2$ and the
coupling constant $\chi$  of our NTS model.

%*********************************
\section{Astronomical tests \label{observ}}

The theoretical model predictions against rotation curve
data of
a set of low surface brightness galaxies taken from Ref.~\refcite{a11,a12}
have recently been tested by Fuchs and Mielke\cite{FM04}.
There the  rotation curves of in total 54 galaxies
have been measured with high resolution. For about half of
them surface photometry is available. For these galaxies not
only kinematical data have been provided,
but also constructed dynamical models of
the galaxies. The observed rotation curves are modelled as
\begin{equation}
  v_\varphi^2(r)= v_{\varphi,\rm bulge}^2(r)+v_{\varphi,\rm disc}^2(r)+
  v_{\varphi,\rm isgas}^2(r)+v_{\varphi,\rm halo}^2(r),
\label{obrotcur}
\end{equation}
where
$v_{\varphi,\rm bulge}$, $v_{\varphi,\rm disc}$, $v_{\varphi,\rm isgas}$, and
$v_{\varphi,\rm halo}$ denote the contributions due to the bulge, the stellar
disc, the interstellar gas, and the dark halo, respectively. The radial
variations of
$v_{\rm c,bulge}(r)$, $v_{\rm c,disc}(r)$, and $v_{\rm c,isgas}(r)$
were derived from the observations, while the normalizations by
the mass--to--light ratios were left as free parameters of the
fits of the mass models to the data. 

{}Fits of the form (\ref{obrotcur}) to observed rotation are notoriously ambiguous.
Thus, for each galaxy several models are provided, one with zero bulge and
disc mass, one model with a `reasonable' mass--to--light ratio of the
bulge and
the disc, and finally a `maximum--disc' model with bulge and disc masses
at the maximum allowed by the data. Furthermore, for each galaxy
two types of dark halo models are tried. One is the cusped NFW density law
and the second is the quasi--isothermal sphere,
whose observational consequences
are recently been reviewed by Bosma\cite{Bo03}. While varying
the disc contribution to the observed rotation curve leads to fits of the
same quality, it is found that the quasi--isothermal sphere
models of the dark halos give significantly better
fits to the data than the cusped NFW density law. Thus the scalar field
model presented here is in this aspect even superior to the cold dark matter
model in its present form.

Prada {\em et al.}\cite{c18} used data from the Sloan digital sky survey 
on satellite galaxies of isolated host galaxies to probe on 100 kpc scale the
{\em outer} halo mass distributions. They find that the line--of--sight
velocity
dispersions of the satellites follow closely the radially declining
velocity
dispersion profile of halo particles in a NFW halo. This implies an outer
mass density distribution of the form $\rho \propto r^{-3}$
which is at variance with
the prediction of the quasi--isothermal sphere (\ref{quasirho}).
In the cold dark matter
model the system of satellite galaxies is assembled during the same
accretion
processes as the dark halo, and Prada {\em et al.}\cite{c18}
assume consistently for the
satellites the same distribution function in phase space as for the halo
particles. In the NTS model, however, the dark halo provides for
the baryonic matter simply a Newtonian force field, 
for which self-gravity has been neglected in zeroth approximation. The distribution function
of satellite galaxies in phase space is thus not specified and can be
modelled
according to the observations, even if the potential trough of the
quasi--isothermal sphere has a shallower profile than a NFW halo.

\begin{figure}
\begin{center}
\epsfxsize=8.0cm
   \leavevmode
      \epsffile{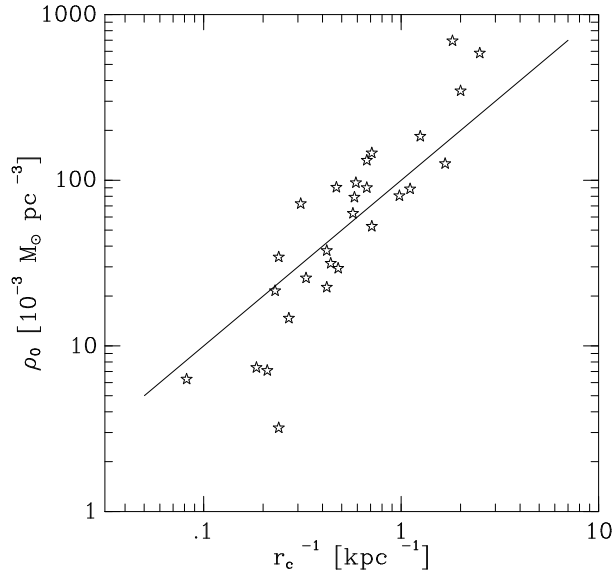}
 \caption{Central densities versus the inverses of the core radii of dark
halos
  of low surface brightness galaxies derived from modelling their rotation
  curves. The halo models are constructed assuming `realistic'
mass--to--light
  ratios for the discs. The solid line is the predicted
  $\rho_0 \propto r_c^{-1}$ relation.}
         \label{imag1}
   \end{center}
   \end{figure}

   \begin{figure}
\begin{center}
\epsfxsize=8.0cm
   \leavevmode
      \epsffile{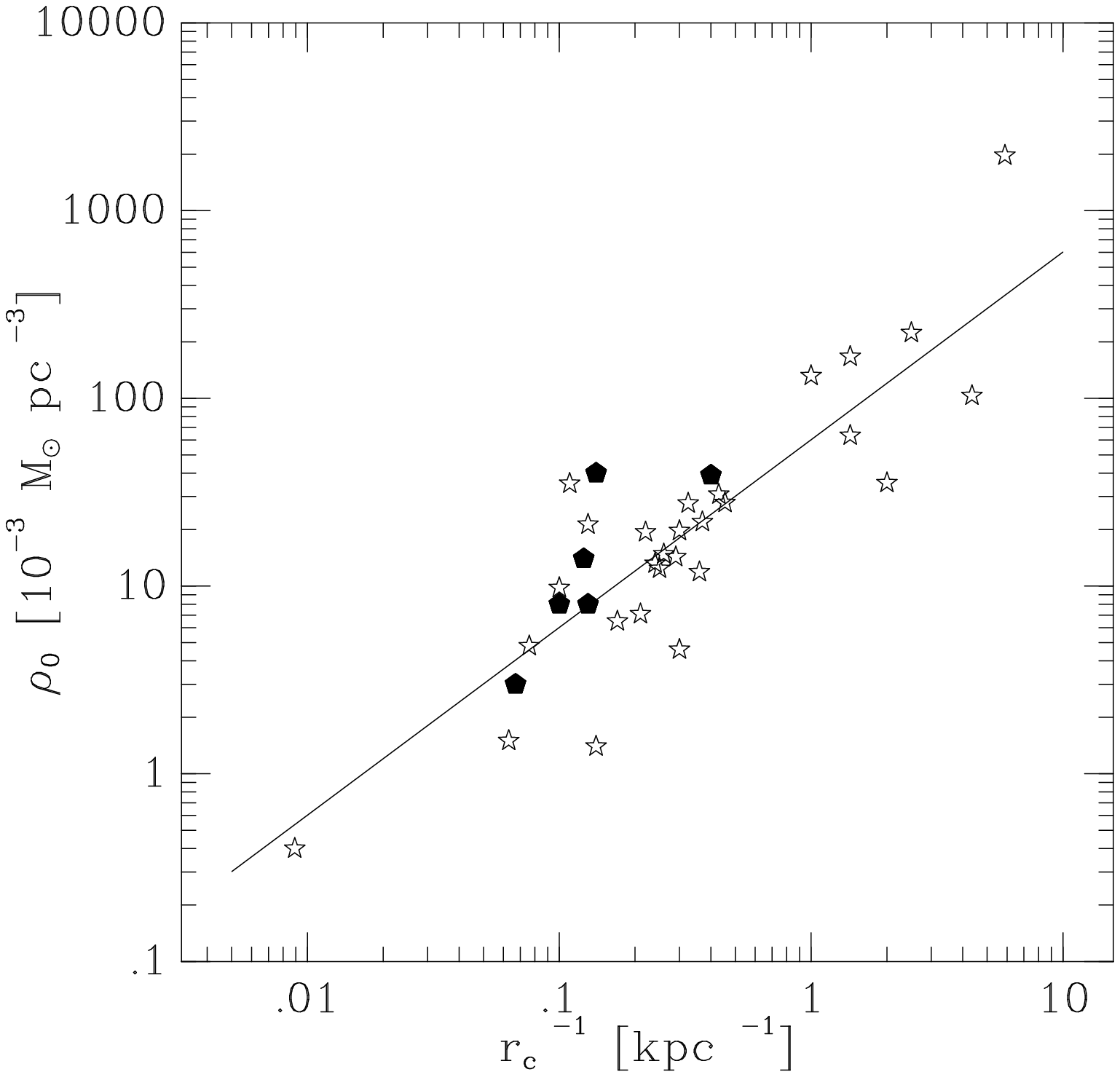}
 \caption{Same as Fig.~\ref{imag1},
  but the halo models are constructed assuming `maximum
  discs'. Open symbols: low surface brightness galaxies, filled symbols:
nearby
  bright galaxies.}
         \label{imag2}
   \end{center}
   \end{figure}

Next we have examined the predicted scaling relation (\ref{scallaw}).
A relation of this type was
found empirically by Salucci and Burkert\cite{SB00},
although this was based on the
universal rotation curve model of Persic {\em et al.}\cite{PS96} and
the, also empirically derived, halo density profile of Burkert\cite{Bu95}.
This density
law resembles the quasi--isothermal sphere in that it has also a
homogeneous core. In Fig.~\ref{imag1}
we show central densities versus the inverses of the core radii
of quasi--isothermal dark halo models constructed in Refs.~\refcite{a11,a12}
assuming for the discs mass--to--light ratios consistent
with current population synthesis models.
Despite some scatter, there is a clear correlation between $\rho_0$ and
$r_c^{-1}$ over several orders of magnitude, $\log(\rho_0) \propto
(1.46\pm 0.55)\log(r_c^{-1})$. Thus the dark halo model data seem to
confirm statistically the scaling relation (\ref{scallaw}).
The scatter in the correlation diagram is probably due to the
near degeneracy of the fits to the observed rotation curves, even if the
disc contributions to the rotation curves are fixed. Moreover, practice shows
that the radial exponential scale lengths of the discs cannot be determined
more precisely than about 20\%. This has a considerable effect for the disc
contributions to the rotation curves (\ref{obrotcur})
and, consequently, for the
dark halo models and might also explain some of the scatter of the
correlation diagram in Fig.~\ref{imag1}.
We believe, however, that this has not changed the general trend.

Using arguments of the density wave theory of galactic
spiral structure, one of us has pointed out, judging from the implied
internal dynamics of the galactic discs, that the discs might be
near to maximum\cite{a18,a15,a19}.
This would imply for the discs of some galaxies mass--to--light
ratios which are significantly higher than in current population synthesis
models of the discs of low surface brightness galaxies. These can be
modified, though, to yield higher mass--to--light ratios.
Therefore we show in Fig.~\ref{imag2} the central densities
versus the inverses of the core radii of the dark halo models, if they are
constructed assuming `maximum discs'. Although the dark halo parameters
are shifted to other values, the linear correlation persists. With
\be
\log(\rho_0)
\propto (1.08 \pm 0.39)\log( r_c^{-1})
\ee
the dark halo models fit nearly
ideally to the predicted scaling relation (\ref{scallaw}).
Included into Fig.~\ref{imag2} are also `maximum disc' dark halo
parameters of nearby bright galaxies\cite{a16,a19}
which fit also well to the scaling law. We conclude from this
discussion that the astronomical tests are rather encouraging for the
scalar field model of dark halos presented here.

%*********************************
\section{General-relativistic corrections\label{outlook}}
In the first part we were focused on a $\Phi^6$ toy model. In order to
advance to more realistic models which include {\em self-gravity}, a full
{\em general-relativistic framework} will be employed which departs from
the coupled Einstein-Klein-Gordon equations
\ba
 G_{\mu \nu }:=  R_{\mu \nu } - \frac{1}{2} g_{\mu \nu } R
             & = & -\kappa T_{\mu \nu } (\Phi ) \; , \label{phi152} \\
      \left (%\Box
              \quabla + \frac {dU}{d\vert  \Phi \vert ^2} \right ) \Phi
             & = & 0 \; . \label{phi153}
\ea
Here $R$ is the curvature scalar, $\kappa = 8\pi G$, the
gravitation constant ($\hbar=c=1$),
$g$ the determinant of the metric $g_{\mu \nu }$, and
$\Phi $ a {\em  complex} scalar field. Moreover,
$
\Box := \left (1/\sqrt {\vert  g\vert  }\right )\,
%\quabla:= \left (1/\sqrt {\vert  g\vert  }\right )\,
 \partial_\mu  \left (\sqrt{\vert  g\vert  } g^{\mu \nu }
 \partial_\nu \right)
$
denotes the generally covariant d'Alembertian.

The stationarity ansatz
\be
\Phi (r,t)=P(r) e^{-i\omega t}
\ee
 describes a spherically symmetric bound
state of the scalar field with frequency $\omega $. Note that
the case of a real scalar field can readily be accommodated
in
our formalism by putting $\omega=0$ in our Ansatz.

For  spherical symmetry of the halo, the line-element reads
\be
ds^2 = e^{\nu (r)} dt^2 - e^{\lambda (r)} dr^2 - r^2 \left(
 d\theta^2 + \sin^2\theta d\varphi^2 \right)  \; ,  \label{met}
\ee
in which the functions $\nu =\nu (r)$ and
$\lambda =\lambda (r)$ depend on the Schwarzschild type radial
coordinate $r$.

Then, the diagonal components of the energy-momentum tensor
$T_\mu{} ^\nu(\Phi) = {\rm diag} \; (\rho , -p_r,$
$-p_\bot, -p_\bot )$ generalize to
\ba
\rho &=& \frac{1}{2} \left ( \omega^2 P^2 e^{-\nu} +P'^2 e^{-\lambda} + U
 \right ) \; , \nonumber  \\
p_{\rm r} &=&  \rho -  U \; , \;  \nonumber  \\
p_\bot &=&  p_r -  P'^2 e^{-\lambda } \; ,
\ea
and the Emden equation (\ref{Emden}) to the radial Klein-Gordon equation
\be
P''(r) + \left ( \frac {\nu' - \lambda'}{2} + \frac {2}{r} \right )
 P'(r) +e^{\lambda}  \left(e^{- \nu }
\omega^2 - \frac{dU(P)}{dP^2} \right) P(r)  = 0  \; . \label{2ska}
\ee

The decisive non-vanishing components of the Einstein equation are  
the `radial' equations
\ba
\nu' + \lambda' & = & \kappa (\rho + p_{\rm r}) r e^\lambda
\; ,  \label{nula}\\
\lambda' & = & \kappa \rho r e^\lambda - \frac {1}{r} (e^\lambda - 1)
\; . \label{la}
\ea
Two further  components are identically fulfilled because of
the contracted Bianchi identity $\nabla^{\mu} T_{\mu}{}^{\nu}\equiv 0$
which is equivalent to the equation
\be
 \frac{d}{dr} p_{\rm r}= -\nu^\prime\left(\rho + p_{\rm r} -\frac{2}{r}(
p_{\rm
 r}-p_\bot)\right)
\ee
of `hydrostatic' equilibrium for
an anisotropic fluid, a generalization\cite{MS96} of the
Tolman-Oppenheimer-Volkoff equation.

The general solution of Eq.~(\ref{la}) is
\be
e^{-\lambda } = 1 - \kappa\frac{M(r)}{r} \rightarrow 1-2v^2_\infty \, ,
\label{lamb}
\ee
where the mass function $M(r)$ is according to (\ref{massf})
only determined by the energy density $\rho$.

According to Eq. (\ref{massfunc}) of the NTS model,
the radial metric component $e^{- \lambda }$ asymptotically approaches
the value
$1-2v^2_\infty<1$. This is not unproblematic: After a redefinition of the
radial coordinate $r \rightarrow \tilde r:= r/\sqrt{1-2v^2_\infty} \; ,$
the asymptotic space has a {\em deficit solid angle}. The area of a sphere
of
radius $r$ is not $4 \pi r^2$, but $4 \pi (1-2v^2_\infty) r^2$;
{\em cf.}~analogous results for global monopoles
and global textures\cite{barr,Sc98}.
 In order to avoid
a linear increase of the mass function which would pose
problems for the asymptotics, GR allows to redefine the radial
coordinate.
In the case of the more {\em realistic} phenomenological
Burkert fit\cite{Bu95} the velocity tends to zero at
spatial infinity, with the consequence that
{\em no} such deficit angle is to be expected.

So far we have considered a solvable model in flat spacetime. However,
when the tangential velocity $v_\varphi= v_\varphi(x)$ is empirically known, we
readily find from (\ref{tanvel}) that
the metric components are in general given by
\ba
e^\nu&=&\exp \left \{ 2 \int v^2_\varphi d\ln x \right \}
  \rightarrow  x^{2 v^2_\infty} + K \, , \label{metnu}
   \\
e^\lambda&=&
  \frac{1 + 2 v^2_\varphi }{1+\kappa p_{\rm r}r^2}
\rightarrow  1 + 2 v^2_\infty
\simeq(1 - 2v^2_\infty )^{-1} \, , \label{metla}
\ea
where $K$ is a constant of integration; in the last line we used
(\ref{lamb}).
The approximation for $e^\lambda$ is valid only for $p_{\rm r}r^2\simeq
0$,
{\em i.e.}~for non-singular radial pressure at the origin or for
sufficiently
fast decreasing pressure at infinity.   Fortunately,
 in  our NTS model, this  condition is satisfied at the
origin as well as at
infinity, {\em cf.}~the asymptotic function (\ref{radpa}).

For our NTS rotation curve (\ref{NTSrot}), we have derived the scalar
field
by setting the metric components to one in zeroth order  Newtonian
approximation.
Then, we calculated the mass function and, consequently, the rotation
curve.
Now, we can do better and determine the metric potentials in first order
approximation. For the lapse function we find
\ba
e^\nu &=& \exp \Biggl \{ \frac{v_\infty^2}{2} \Bigl [
 \ln \Bigl \{ x^7 \sqrt{1+x^2} \left ( \frac{1+A^2x^2}{A^2x^2} \right
)^{7/2}
     \Bigr \} \nonumber\\
 &\quad& \qquad\qquad - \frac{1}{1+x^2}+ \frac{7A}{x}\arctan (Ax)
 \Bigr ] \Biggr \} \; .
\ea
The shift function $e^\lambda$ is influenced by both,  the rotation
velocity
(\ref{NTSrot}) and  the radial scalar field pressure (\ref{radp}). Therefore, a full account of self-gravity 
seems to demand  numerical methods.

A more ambitious approach is to {\em reconstruct}
a viable scalar potential  $U(\vert \Phi \vert )$ from the empirical
rotation curves on the basis of the Einstein equations,
similarly as in the case of inflation\cite{MS95}.

As a starting point, we may use
the empirical Burkert profile\cite{Bu95} which can be expressed in the following 
close analytical form: 
\ba
 v^2_{\varphi{\rm B}}/v^2_\infty = \frac{1}{2x}\left\{\ln[(1+x)^2(1+x^2)] -
2\arctan(x)\right\}&\simeq& 1 -\frac{\arctan(x)}{x} \nonumber\\
&\rightarrow& 2\frac{\ln x}{x}\,.  \label{Bur}
\ea
It has  a maximum at $x=3.3$ in dimensionless units $x=r/r_{\rm c}$ and amounts
at spatial infinity  to a {\em  logarithmic modification} of the
Kepler law. Then,
we find for the lapse function via {\sc Reduce}
{\small
\be
 e^\nu = \exp \left \{ \frac{v^2_\infty}{x}\left[x
\ln\frac{1+x^2}{(1+x)^2} +
  2(1+x)\arctan(x) -
 \ln\left(1+2x +2x^2+2x^3+x^4\right)\right] \right \} \; .
\ee}
Assuming a negligible radial pressure term, i.e. $p_{\rm r}r^2\simeq 0$,
the shift function $e^\lambda = 1 + 2 v^2_\varphi $ is simply determined
by the Burkett rotation velocity (\ref{Bur}).

For this Burkert fit, we have  attempted to {\em reconstruct}\cite{MS02} a
self-interaction
potential $U(\Phi) =\rho -p_{\rm r}$ from the astronomical data,
 by resolving the radial Einstein
equations (\ref{la}) for $\rho$ and $p_{\rm r}$ with the general result
\be
U = \rho -p_{\rm r}=\frac{1}{\kappa r e^\lambda}\left[\lambda^\prime
 -\nu^\prime + \frac{2}{r}(e^\lambda -1)\right] \; . \label{potre}
\ee

Accordingly, $U$ is  positive when
\be
d\ln v_\varphi/d\ln x + v^2_\varphi +1/2 \geq 0
\ee
 holds, which is satisfied by the Burkert fit.
However, since the radial dependence of the scalar field $\Phi(x)$ is still
unknown, a numerical approach is needed in order  to  reconstruct
the potential $U(\Phi)$ explicitly. 

In view of the occurrence of scalar fields, like the dilaton or
Kalb-Ramond axion in effective superstring models leading to
{\em axidilaton stars}\cite{MS01}, the concordance of
theory and observations in a viable self-interaction potential
deserves further study.

%****************************
\section{Gravitational lensing of dark halos \label{lensing}}

Another promising approach to study the dark halos of galaxies is weak 
{\em gravitational lensing}, since 
it can be measured out to large projected distances from the lense.
Hence, gravitational lensing of a
{\em transparent} spherically symmetric NTS dark matter halo will be 
 compared here with 
the scenario of a mini-BS lens\cite{SM99,DS00}.
We assume that the lens interior
is almost empty of baryonic matter, such that deflected
photons can travel freely through the space between the halo stars
which embody only ``small'' disturbances\cite{Mi02}.

The deflection angle is then given by
\be
\hat{\alpha}(r_0) = 2 \int\limits_{r_0}^\infty
\frac{be^{\lambda /2}}{r\sqrt{r^2 e^{-\nu} - b^2}} dr  - \pi \; ,
\label{alpha}
\ee
where $b=r_0 \exp (-\nu(r_0))$ is the impact parameter
and $r_0$ denotes the closest distance between a light ray and
the center of our halo.
The lens equation {\em for small deflection angles} can be expressed as
\be
\sin\left(\theta-\beta\right)
 = \frac{D_{\rm ls}}{D_{\rm os}} \sin\hat{\alpha} \, ,
 \label{lens1}
\ee
where $D_{\rm ls}$ and $D_{\rm os}$ are the distances from the lens
(deflector) to the source and from the observer to the source,
respectively.
The true angular position of the source is denoted by $\beta$,
whereas $\theta$ stands for the image positions.
One usually defines the {\em reduced deflection angle} to be
$
\alpha \equiv \theta - \beta = \sin^{-1}\left(D_{\rm ls}
\sin\hat{\alpha} /D_{\rm os} \right) \, .
$
However, equation (\ref{lens1}) relies on substitution of the
distance from the source to the point of minimal approach by the
distance from the lens to the source $D_{\rm ls}$.
{\em For large deflection angles} the distance $D_{\rm ls}$
cannot be considered a constant but it is a function of the deflection
angle so that the form of the lens equation changes\cite{DS00} into
\begin{equation}
\sin{\alpha} = \frac{D_{\rm ls}}{D_{\rm os}} \cos{\theta}
\cos \left[ \mbox{arsin}
\left ( \frac{D_{\rm os}}{D_{\rm ls}} \sin (\theta - \alpha )
  \right) \right]
\left[\tan{\theta} + \tan(\hat{\alpha} - \theta)
\right] \, .  \label{lens2}
\end{equation}
The reduced deflection angle can be kept defined as
$\alpha \equiv \theta - \beta$; {\em cf.}~Ref.~\refcite{Pe03}
with more details on the lens equation.

Since our dark matter halo, the NTS soliton, has only a weak
gravitational field, we can apply (\ref{lens1}) together with
(\ref{alpha}). The deflection angle is determined by
the gravitational potentials (\ref{metnu}), (\ref{metla})
which are functions of the NTS rotation curve (\ref{NTSrot})
and the radial pressure (\ref{radp}). As an imput, we use the best fit
value\cite{MSP02} of $A=0.805$ to the Burkert profile (\ref{Bur}).
The asymptotic rotation velocity is taken to be $v_\infty = c/1000$.

\begin{figure}[t]
\begin{center}
\leavevmode\epsfysize=6.5cm
      \epsffile{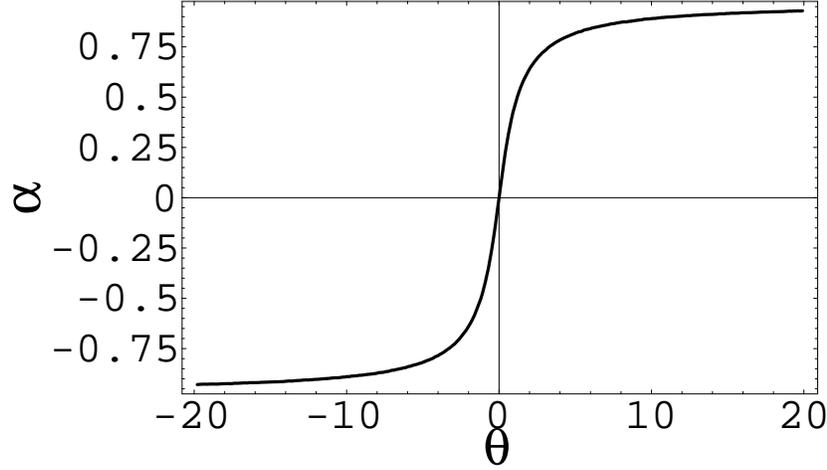}
 \caption{Reduced deflection angle for the NTS dark matter halo.}
         \label{imag3}
  \end{center}
   \end{figure}

Numerical computations of the reduced deflection angle for the NTS
halo were performed by assuming that the halo lens is half-way between the
observer and the source, i.e.~$D_{\rm ls}/D_{\rm os}=1/2$.
The result is shown in Fig. \ref{imag3}
where $\alpha$ and $\theta$ are measured in arcsecond, but in general
the units of $\theta$ depends on the oberserver-lens distance
(further details, see Refs.~\refcite{DS00,ST00}).
We recognize prominent features of the NTS lens 
due to the linear increasing mass (\ref{massfunc}).
There exists a limiting value of the reduced deflection angle
of about $\alpha =0.93$ arcsecond. Hence, if the true angular
position of the source $\beta$ increases, nevertheless
the halo lens deflects the light with a constant value.
This behavior would change if the mass of the halo becomes finite.

We can compare this result with the strong gravitational source
of a BS for different potentials $U$, when 
the BS lens is again half-way between the
observer and the source\cite{DS00,ST00}.
Observable differences are found depending on the choice of the
self-interaction. In the case of a simple mass term 
$U(\vert  \Phi \vert ) = m^2 \vert  \Phi \vert ^2$,
the largest possible value of $\alpha$ is
23.03 degrees with an image at about $\theta = n \times 2.88$
arcsec where $n=n(D_{\rm ol},\omega)$ is the distance factor which is a
function of the distance from the observer to the lens and the
scalar field frequency the inverse of which is associated with the
BS radius. For non-relativistic BS approximations, smaller angles will
occur. The angle $\theta$ of the image position can have very
different orders of magnitude, depending on $n$.
For example, $n=1$ fixes $\theta$ to be measured in arcsec.
Under the assumption that the BS mass is 10$^{10}$ $M_{\odot}$,
the distance $D_{\rm ol}$ is about 100 pc.
If the distance factor is $n=10^{-3}$,
then $\theta $ is measured in milli-arcsec and the
BS-lens is at about 100 kpc from the observer.

In comparison, halos modelled by the Burkert profile will not
produce strong lensing as well\cite{PF03}.
Differences of our NTS model and the Burkert fit can be
traced back to the fact that the metric corresponding to an
asymptotically constant velocity $v_\infty$ in general
exhibits a {\em deficit angle} similarly as a cosmic string\cite{Sr87}.
The linear mass density peaks at the location of the string,
thereby producing a strong lensing.

The properties of dark matter halos inferred from weak lensing\cite{HY03}  
provides a strong support for 
the existence of dark matter, wheras alternative theories of
gravity such as MOND can almost be excluded.

%*********************************
\section*{Acknowledgments}

We would like to thank Fjodor Kusmartsev, Humberto Peralta and Remo Ruffini
for helpful discussions and comments.
One of us (F.E.S.) acknowledges
research support provided by a personal fellowship.
Moreover, (E.W.M.) acknowledges the support of SNI and
thanks  Noelia,  Markus G\'erard Erik, and Miryam Sophie Naomi
for encouragement.

%*********************************
%\section*{References}

{}%for Latex
%\end{references} %for REVTEX
\end{document}